\documentclass[manuscript]{emulateapj}

\usepackage{apjfonts}

\slugcomment{to appear in the Astrophysical Journal}

\shorttitle{Forced Novae} 
\shortauthors{Hachisu  et al.}

\makeatletter
\AtBeginDocument{\@ifpackageloaded{natbib}{\ifNAT@numbers\if@filesw\immediate\write\@auxout{\string\global\string\NAT@numberstrue}\fi\fi}{}}
\makeatother
\begin{document}


\title{Shortest Recurrence Periods of Forced Novae}


\author{Izumi Hachisu}
\affil{Department of Earth Science and Astronomy, College of Arts and
Sciences,  University of Tokyo, 3-8-1 Komaba, Meguro-ku,
Tokyo 153-8902, Japan}
\email{hachisu@ea.c.u-tokyo.ac.jp}

\author{Hideyuki Saio}
\affil{Astronomical Institute, Graduate School of Science,
    Tohoku University, Sendai 980-8578, Japan}

\and

\author{Mariko Kato} 
\affil{Department of Astronomy, Keio University, Hiyoshi, Yokohama
  223-8521, Japan}



\begin{abstract} 
We revisit hydrogen shell burning on white dwarfs (WDs) with 
higher mass accretion rates than the stability limit, 
$\dot M_{\rm stable}$, above which  hydrogen burning is stable. 
Novae occur with mass accretion rates below the limit. 
For an accretion rate \textgreater\ $\dot M_{\rm stable}$,
a first hydrogen shell flash occurs followed by steady nuclear burning,
so the shell burning will not be quenched 
as long as the WD continuously accretes matter.  
On the basis of this picture, 
some persistent supersoft X-ray sources can be explained by 
binary models with high accretion rates.
In some recent studies, however, 
the claim has been made that no steady hydrogen 
shell burning exists even for accretion rates \textgreater\ 
$\dot M_{\rm stable}$. 
We demonstrate that, in such cases, repetitive flashes occurred
because mass accretion was artificially controlled.
If we stop mass accretion during the outburst, no new nuclear fuel
is supplied, so the shell burning will eventually stop.  If we resume
mass accretion after some time, the next outburst eventually occurs. 
In this way, we can design the duration of outburst and interpulse time
with manipulated mass accretion.  We call such a controlled nova
a ``forced nova.''
These forced novae, if they exist, could have much shorter recurrence
periods than ``natural novae.''  We have obtained the shortest recurrence
periods for forced novae for various WD masses.
Based on the results, we revisit WD masses of some recurrent novae
including T~Pyx.
\end{abstract}

\keywords{ nova, cataclysmic variables -- stars: individual (T~Pyx)
-- white dwarfs  -- X-rays: binaries 
}

\section{Introduction}
\label{sec_introduction}
A classical nova is a thermonuclear runaway (unstable hydrogen shell
flash) event on a mass-accreting white dwarf (WD), which occurs
if the mass accretion rate $\dot M_{\rm acc}$ is smaller than the
stability limit $\dot M_{\rm stable}$ corresponding to the WD mass.
Many theoretical works on hydrogen shell flashes have been published.
In general, shorter decay times of novae are obtained for more massive
WDs \citep[e.g.,][]{hac06kb, hac10k, hac14k, hac15k, hac16k}, and
shorter recurrence periods correspond to more massive WDs
with higher mass accretion rates 
\citep[e.g.,][]{pri95, wol13, wol13b, kat14shn}.

If the mass accretion rate exceeds $\dot M_{\rm stable}$, nuclear burning
is stable and no repeating shell flashes occur.  This stability
has long been studied in analytical and numerical works
\citep{pac78,sie75, sie80, sio79, ibe82, nom07, she07, wol13, kat14shn}.
The physical reason of stabilization was also presented by
many authors \citep{sug78, fuj82, yoo04, she07, she08}.
The border between the stable and unstable mass accretion rates is known
as the stability line, i.e., $\dot M_{\rm stable}$,
in the diagram of accretion rate versus WD mass \citep[e.g.,][]{nom82}.

There is another critical accretion rate, above which the 
optically thick winds occur \citep{hkn96, hac01kb}.   
A hydrogen-rich envelope of a WD blows optically thick winds 
\citep{kat94h}, instead of slowly expanding to a red-giant size
\citep{nom82}.   
This critical mass accretion rate is dubbed as $\dot M_{\rm cr}$, which 
is about twice  $\dot M_{\rm stable}$ \citep[e.g.,][]{kat14shn}.


\begin{figure}
\epsscale{1.15}
\plotone{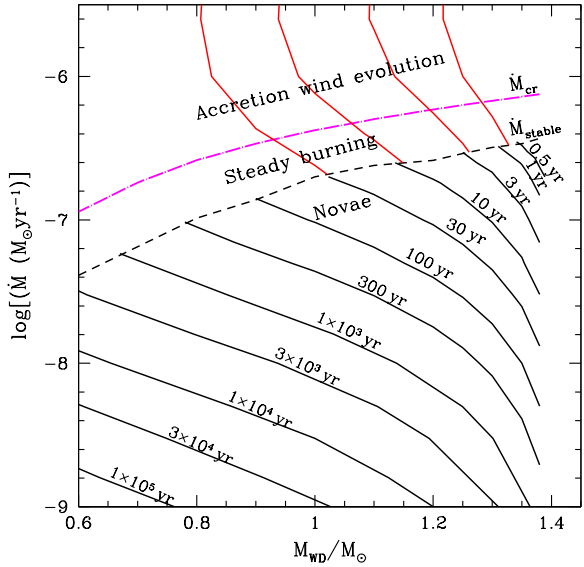}
\caption{
Response of a WD envelope to mass accretion in the WD mass versus 
mass accretion rate.  
Above the stability line denoted by the black dashed line
labeled $\dot M_{\rm stable}$, hydrogen shell burning is stable.
Below the stability line (labeled ``Novae''), shell flashes repeat with
the recurrence period denoted by black solid lines.
Above the critical line denoted by the magenta dash-dotted line,
optically thick winds blow (labeled ``Accretion wind evolution'').  
The data are taken from Figure 6 of \citet{kat14shn} for natural novae.
We added red lines of recurrence periods of 1, 3, 10, and 30 yr
for forced novae.  See text for more details.
\label{PrecWDmass}}
\end{figure}

\begin{figure}
\epsscale{1.0}
\plotone{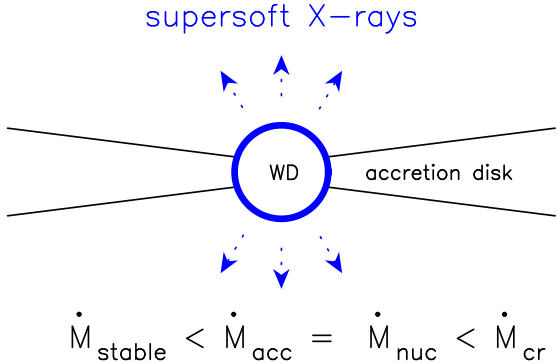}
\caption{
Schematic configuration of steady hydrogen shell burning 
with no wind mass loss.  The WD accretes matter from an accretion disk 
and burns hydrogen at the same rate as the accretion.
No nova outburst occurs if the accretion rate is kept  above
the stability line.  Such a configuration is often referred to as 
a persistent supersoft X-ray source.
\label{steadyaccretion}}
\end{figure}

\begin{figure}
\epsscale{1.0}
\plotone{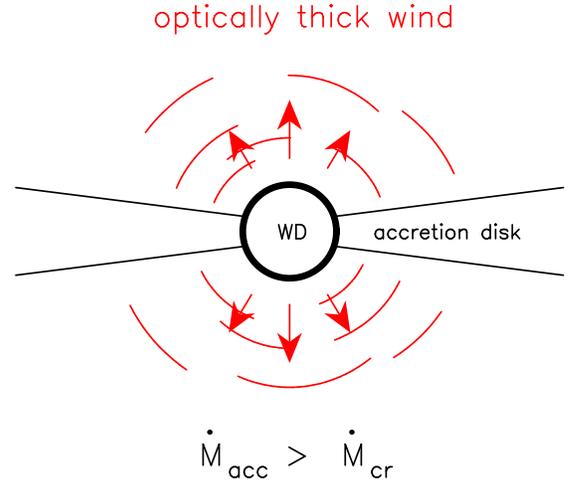}
\caption{
Schematic configuration of  accretion wind evolution. 
Hydrogen steadily burns on the WD.  The WD accretes matter
from a disk and blows excess matter into the wind. 
See text for more details.
\label{accretionwind}}
\end{figure}

In summary, when the mass accretion rate $\dot M_{\rm acc}$
is less than the stability 
line, $\dot M_{\rm acc} < \dot M_{\rm stable}$, the nuclear burning
is unstable and the WD undergoes a number of nova outbursts. 
If the mass accretion rate is between the two critical values, 
i.e., $\dot M_{\rm stable} < \dot M_{\rm acc} <  \dot M_{\rm cr}$,
the hydrogen shell burning is stable (see Figure \ref{PrecWDmass}). 
The accreted hydrogen-rich matter burns steadily on the WD
at the same rate as the mass accretion rate.
We suppose a binary configuration as illustrated
in Figure \ref{steadyaccretion}; that is, 
the WD accretes matter from an accretion disk.  
Such a configuration has been used to explain
supersoft X-ray sources (SSSs) \citep[e.g.,][]{vdh92,sch97}. 
If the mass accretion rate is larger than $\dot M_{\rm cr}$, 
i.e., $ \dot M_{\rm acc} >  \dot M_{\rm cr}$, 
optically thick winds blow from the surface of the envelope, and, 
at the same time, the WD accretes matter from the disk 
(see Figure \ref{accretionwind}). 
Hydrogen burning is stable and the accreted matter is partly  
burned into helium and accumulated onto the WD. 
The excess matter is lost by winds, with the mass-loss rate of
$\dot M_{\rm wind} = \dot M_{\rm acc} - \dot M_{\rm nuc}$.
Such a state was dubbed ``accretion wind evolution'' \citep{hac01kb} 
and has been regarded as the state of luminous SSSs
\citep[also see][]{hac03RXJ,hac03VSge}.  
The most recent versions of the stability line and critical line
for winds appeared in Figure 5 of \citet{kat14shn}.  


\begin{figure}
\epsscale{1.15}
\plotone{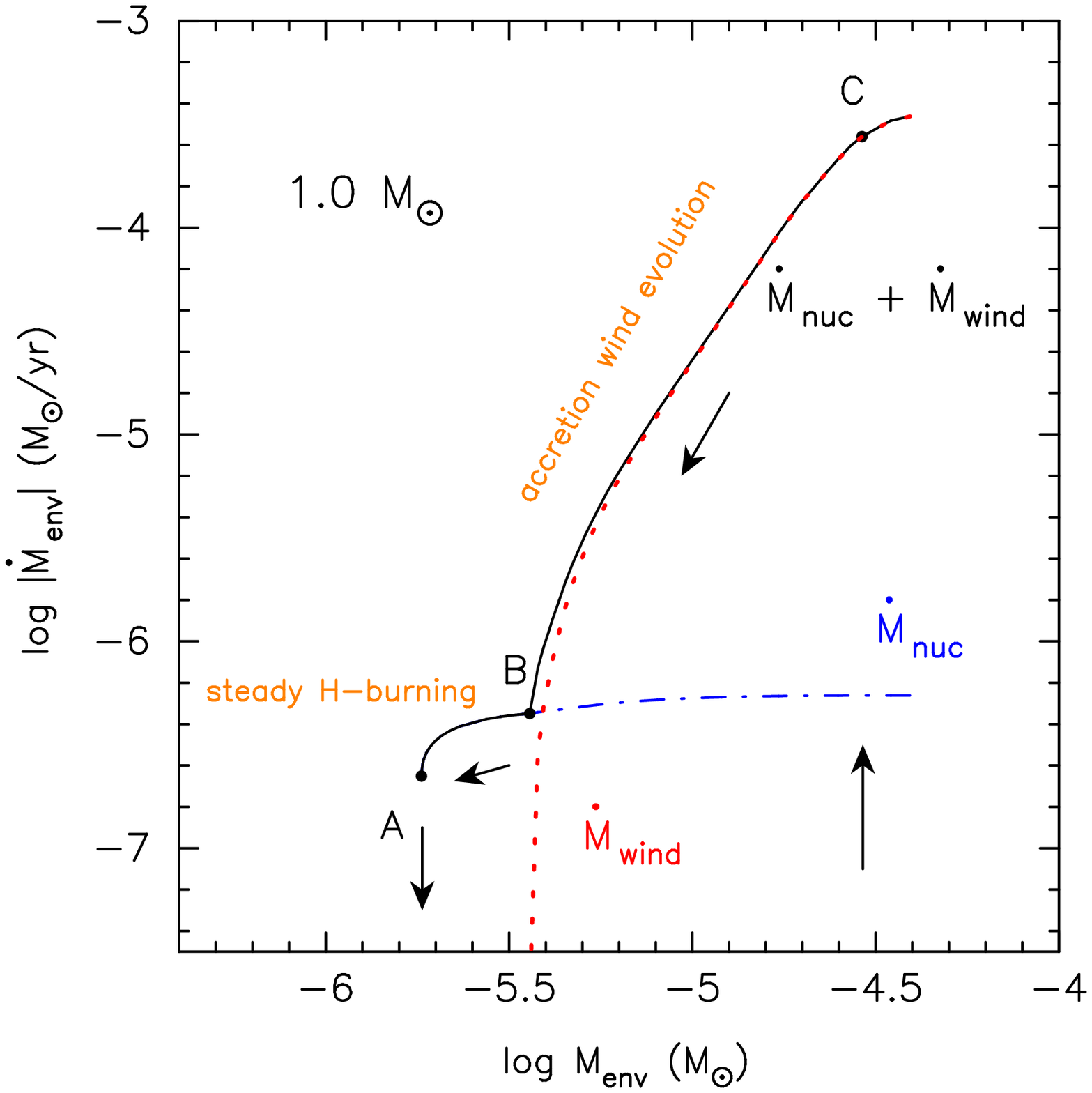}
\caption{
Mass decreasing rate of the hydrogen-rich envelope
($-\dot M_{\rm env}=\dot M_{\rm wind} + \dot M_{\rm wind}$)
versus envelope mass ($M_{\rm env}$) on a $1.0~M_\sun$ WD.
The data are taken from the calculation of \citet{kat94h}
for a $1.0~M_\sun$ WD with the envelope of solar composition.   
We assume steady state for the hydrogen-rich envelope on the WD.
Hydrogen shell burning is stable when the envelope mass $M_{\rm env}$
is larger than $M_{\rm env,min}\equiv M_{\rm env}$(A) at point A,
i.e., $M_{\rm env} > M_{\rm env,min}$.
Optically thick winds blow when the envelope mass $M_{\rm env}$
is larger than $M_{\rm env,cr}\equiv M_{\rm env}$(B) at point B, i.e.,
$M_{\rm env} > M_{\rm env,cr}$.
The mass decreasing rates are $\dot M_{\rm stable}= - \dot M_{\rm env}$(A)
at point A, and $\dot M_{\rm cr}= - \dot M_{\rm env}$(B) at point B.
For an envelope mass of the ignition mass $M_{\rm ig}$,
the nova reaches point C, where $M_{\rm ig}=M_{\rm env}$(C),
at optical maximum and then evolves
down to point A through B along the steady-state sequence
in the direction of arrows.  
\label{dmdtdM}}
\end{figure}

To understand the nova cycle based on the above picture,
we plot variations of envelope mass during one cycle of nova outburst
in Figure \ref{dmdtdM}, a model of steady-state envelope solutions
on a $1.0~M_\sun$ WD \citep{kat94h, hac01kb}.  
When the envelope mass reaches a critical value,
hydrogen ignites to trigger a nova outburst 
(denoted by the upward arrow).
The envelope expands to reach optical maximum (point C). 
The optically thick winds are accelerated so that the envelope mass
decreases by wind mass loss ($\dot M_{\rm wind}$)
and hydrogen burning ($\dot M_{\rm nuc}$).
The envelope evolves down along with the black line
with the photospheric temperature increasing. 
When the photospheric temperature reaches 
the OPAL opacity's peak ($\log T$ (K) $\sim 5.2$), 
the optically thick winds cease (point B). 
We define this point as $M_{\rm env,cr}\equiv M_{\rm env}$(B), 
$\dot M_{\rm cr}=\dot M_{\rm nuc}$(B), and $\dot M_{\rm wind}$(B)$=0$,
above which we have optically thick wind solutions. 
After that, the envelope mass still decreases  
because of nuclear burning. 
When it reaches point A, hydrogen nuclear burning diminishes
and the WD cools down.  There exist equilibrium hydrogen-burning 
envelope solutions below 
point A but they are unstable \citep[see Figure 1 of][]{kat14shn}.
We define this envelope mass as $M_{\rm env,min}\equiv M_{\rm env}$(A)
and $\dot M_{\rm stable}=\dot M_{\rm nuc}$(A).

If the ignition mass is smaller than that at point B in Figure \ref{dmdtdM},
i.e., $M_{\rm ig} < M_{\rm env,cr}= M_{\rm env}$(B), 
the nova reaches a point on the sequence between points A and B 
where no optically thick winds are accelerated. 
The envelope expands only slightly and the effective temperature does not 
decrease much, so it would be bright in the supersoft X-ray or UV band.
It should be addressed that, if the accretion rate is high enough to be
close to $\dot M_{\rm stable}$, the accretion to the WD during hydrogen
shell burning makes the effective decreasing rate of the envelope
mass smaller and, as a result, the nova-on (SSS) phase
becomes significantly longer.

Point A for various WD masses describe the stability line 
$\dot M_{\rm stable}$ in Figure \ref{PrecWDmass} while 
point B for each WD mass corresponds to the line of the critical mass
accretion rate $\dot M_{\rm cr}$ for winds.
If the mass-accretion rate $\dot M_{\rm acc}$ is between 
$\dot M_{\rm stable}$ and $\dot M_{\rm cr}$, i.e., 
$\dot M_{\rm stable} < \dot M_{\rm acc} < \dot M_{\rm cr}$, 
the hydrogen-rich envelope stays somewhere
between A and B in Figure \ref{dmdtdM} and keeps
a steady-state of $\dot M_{\rm nuc}= \dot M_{\rm acc}$ without
wind mass-loss ($\dot M_{\rm wind}=0$).    
When the mass-accretion rate is larger than $\dot M_{\rm cr}$, 
i.e., $\dot M_{\rm acc} > \dot M_{\rm cr}$, 
the envelope stays somewhere
between B and C (or above C) in Figure \ref{dmdtdM} and keeps
a steady-state of $\dot M_{\rm wind} + \dot M_{\rm nuc}= \dot M_{\rm acc}
> \dot M_{\rm cr}=\dot M_{\rm nuc}$(B).    

These basic properties of mass-accreting WDs play essential roles
in the evolution of binaries toward type Ia supernovae (SNe Ia).
If the mass of a mass-accreting carbon-oxygen WD approaches
the Chandrasekhar mass, carbon ignites to trigger an SN Ia explosion
\citep{nom82}.  
The modern single degenerate (SD) scenario is based on the binary
evolution theory 
in which both steady hydrogen shell burning (Figure \ref{steadyaccretion}) 
and accretion wind evolution (Figure \ref{accretionwind}) 
phases are taken into account in the evolutional paths to SNe Ia
\citep[e.g.,][]{hkn96, hknu99, hkn99, hkn10, hksn12, hkn12, 
li97, lan00, han04}.  


Some groups have recently claimed,  however, that there is no steady hydrogen 
shell burning on mass-accreting WDs even above the stability line
\citep[e.g.,][]{sta12bal, ida13}.
\citet{kat14shn} elucidated the reason why they did not have steady
hydrogen shell burning.  Kato et al. showed two different evolutions
of mass accretion.  They started the mass accretion onto a WD 
(with no hydrogen burning) at a rate higher than the stability line
and obtain a first shell flash.  They stopped the mass accretion during
the mass-loss phase.  After some time elapsed 
(but with hydrogen shell burning still occurring), they restarted the mass
accretion and obtained continuous shell burning (steady-state burning).
However, they obtained repeated shell flashes 
if they did not start the mass accretion until
hydrogen shell burning began to decay. 
These flashes are obtained only when they controlled the
on/off epochs of mass accretion. 
Thus, one can obtain shell flashes above the stability
line if one can control the on/off epochs of mass accretion.
After Kato et al.'s (2014) paper was published, 
\citet{hil15} further claimed that they did not find steady
burning simply because their numerical code produced a series
of shell flashes.  Based on their time-dependent calculations,
\citet{hil15} concluded that steady hydrogen shell burning,
on which Hachisu \& Kato's accretion wind evolution model depends,
does not occur even above the stability line.

Motivated by such confusion, we study shell flashes 
above the stability line in a more systematic way. 
Here we call them ``forced novae'' after forced oscillation in physics. 
The forced novae occur only if we control the on/off epochs of
mass accretion.  In contrast, shell flashes naturally
occur {\it below the stability line}, so we call them ``natural novae.''

Forced novae may occur if the mass accretion is controlled by some
mechanism in the binary system.  For example, if the accretion disk
is destroyed or if the mass transfer from the 
companion stops, the WD has some period without mass accretion until the 
accretion condition is recovered.  For classical novae, which are
considered to be systems below the stability line,
this situation has already been discussed
as ``hibernation'' \citep[e.g.,][]{sha86}. 

Forced novae could have much shorter recurrence periods than those
of natural novae.
We obtain the shortest recurrence period of forced novae
and compare them with the recurrence period of natural novae.
These recurrence periods give an important constraint
on the WD masses of recurrent novae.
For example, for the Galactic recurrent nova U~Sco, whose recurrence period
is $t_{\rm rec} \sim 8$--12~yr, the WD mass is constrained to be 
$M_{\rm WD} > 1.15~M_\sun$ from the recurrence periods of
natural novae \citep[see Figure 6 of][]{kat14shn}.
In the same way, the minimum periods of forced novae can be used 
to constrain the WD mass. 
Using this constraint, we discuss the WD mass in T Pyx.

We organize the present paper as follows.  
In Section \ref{section_time_depend} we
describe various forced nova evolutions using a time-dependent
evolution code and  in Section \ref{sec_Pmin} 
we obtain the shortest recurrence periods of
forced novae.  We apply our results to
various recurrent novae in Section \ref{sec_discussion} and 
constrain the WD mass of T~Pyx by the shortest
recurrence periods of forced novae.
Finally we summarize our results in Section \ref{sec_conclusion}.


\begin{figure*}
\epsscale{0.8}
\plotone{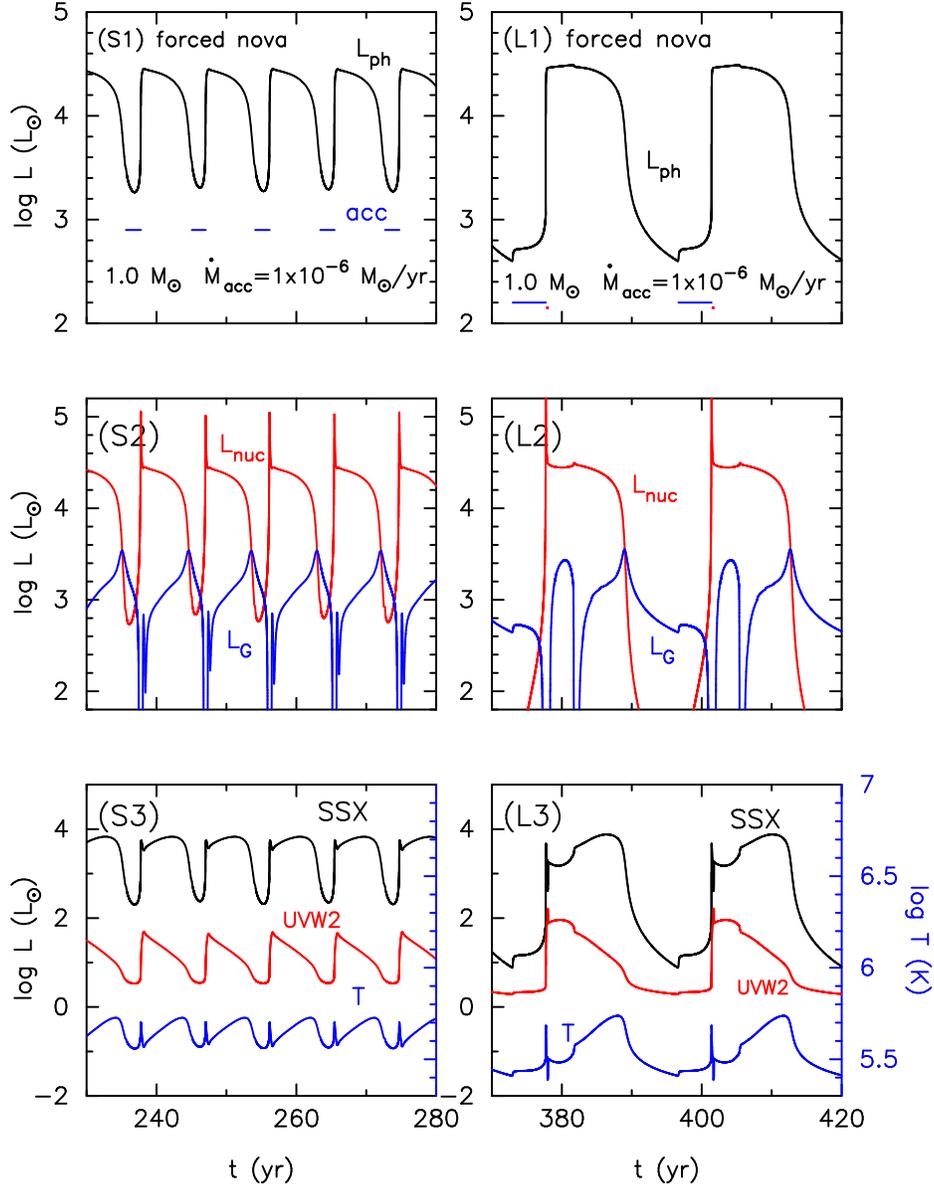}
\caption{
Time-dependent calculations of shell flashes for forced novae
on a 1.0 $M_\sun$ WD with a mass
accretion rate of $1 \times 10^{-6}~M_\sun$ yr$^{-1}$. 
Left panels (S1--S3, where ``S'' stands for short cycle):
The last five cycles of our forced nova calculation with immediate 
resumption of accretion.   
(S1) Photospheric luminosity $L_{\rm ph}$. The accretion phase
is indicated by the short horizontal blue lines labeled ``acc.''
The recurrence period is 9.2 yr.  The on/off epochs of mass accretion
are controlled to match the recurrence period with that of Idan et al. (2013). 
(S2) Nuclear luminosity $L_{\rm nuc}$ and gravitational energy
release $L_{\rm G}$.  The negative values of $L_{\rm G}$ are cut off
by the lower bound of this figure. The photospheric luminosity 
in the interpulse phase is mainly supplied by $L_{\rm G}$.
(S3) Photospheric temperature $T_{\rm ph}$, 
flux of supersoft X-rays (0.2--1 keV), and UV flux (1120--2640 \AA)
corresponding to the {\it Swift} $UVW2$ band.
These luminosities are calculated from a blackbody assumption of
the photospheric temperature and luminosity. 
Right panels (L1--L3, where ``L'' stands for  long cycle):
The same WD mass and mass accretion rate as those in the left
panels, but with delayed on time of mass accretion. 
The recurrence period is 23.7 yr.  See text for more details.
\label{L.m10.1E-6.two}}
\end{figure*}


\begin{figure}
\epsscale{1.15}
\plotone{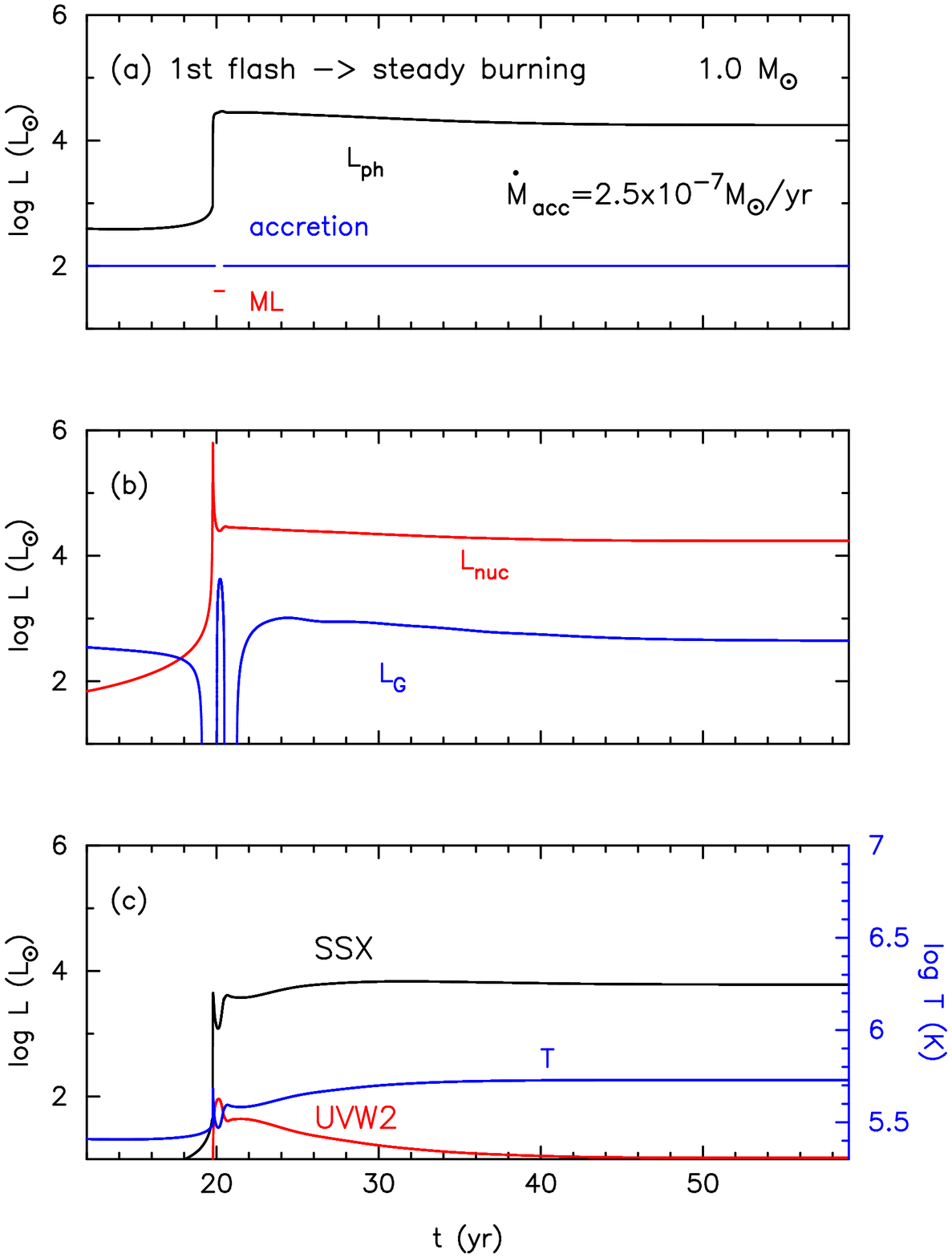}
\caption{
Same as   Figure \ref{L.m10.1E-6.two}, but for
an immediate on time of mass accretion.  The mass accretion rate
is assumed to be $2.5\times 10^{-7}~M_\sun$~yr$^{-1}$.
The WD undergoes a first shell flash and then settles down to
steady-state hydrogen shell burning.
\label{L.m10.stburn}}
\end{figure}

\section{Time-Dependent Evolutions of Forced Novae}
\label{section_time_depend}
In principle, no nova outbursts occur above the stability line, i.e., 
$\dot M_{\rm acc} > \dot M_{\rm stable}$, in Figure \ref{PrecWDmass}
because nuclear burning is stable.  However, there are two exceptions. 
One is the ``first shell flash'' (the first nova outburst) that occurs
only once on the WD after it begins mass accretion.  
Suppose that a naked WD begins accretion.
Hydrogen-rich matter accumulates on the WD.
When it reaches the ignition mass, a hydrogen shell flash inevitably
occurs.  This is the first shell flash. 
If the mass accretion stays high as
$\dot M_{\rm acc} > \dot M_{\rm stable}$, 
hydrogen shell burning becomes stable and never stops; i.e., 
the WD keeps bright.
Thus, the first shell flash is the first and the last nova outburst
for the WD.  It never repeats nova outbursts.
Then the binary configuration is either that in 
Figure \ref{steadyaccretion} or that in Figure \ref{accretionwind}. 
The light curve for such a case has already been reported, e.g., 
in Figure 7(a) of \citet{kat14shn}.

The other case is a forced nova.
If we stop mass accretion at some epoch during shell flashes,
the WD cannot keep steady hydrogen burning
because of the shortage of nuclear fuel. 
The hydrogen shell burning eventually stops. 
Then, we resume mass accretion and have the next outburst. 
In this manner, we have successive nova outbursts
by manipulating mass accretion on and off.  The quiescent phase is 
determined by the epoch when we switch mass accretion on.  
In this way, we can freely design a nova outburst
for an accretion rate above the stability line, that is,
$\dot M_{\rm acc} > \dot M_{\rm stable}$,
in Figure \ref{PrecWDmass}.   
This is the forced nova.

\subsection{Numerical method}
\label{numerical_method}
Evolution models of mass-accreting WDs were
calculated by using the same Henyey-type code as in \citet{kat14shn}. 
This code implements zoning based on the fractional mass ($q\equiv M_r/M$)
in outer layers of the hydrogen-rich envelope, whereas zoning is based
on  $M_r$ in the rest of the interior including nuclear burning layers. 
In the outer layers the gravitational energy release per unit mass,
$\epsilon_{\rm g}$, is 
calculated as \citep[e.g.,][]{neo77}
\begin{equation}
\epsilon_{\rm g} \equiv -T\left(\partial s\over\partial t\right)_{M_r}
= -T\left(\partial s\over\partial t\right)_q + T{d\ln M\over dt}
\left(\partial s\over\partial\ln q\right)_t, 
\end{equation}
where $M$ is the total mass of the WD, $M_r$ is the mass within
the radius $r$, and $T$ and $s$ are the temperature and specific
entropy, respectively.  This zoning scheme with an off-center
differencing for the equation of energy conservation \citep{sug70}
works well for rapid evolution with mass accretion or decretion.
Accretion energy outside the photosphere is not included. 

The chemical composition of the accreting matter and initial
hydrogen-rich envelope
is assumed to be $X=0.7, ~Y=0.28$, and $Z=0.02$.
Neither convective overshooting nor diffusion processes of nuclei are
included; thus, no WD material is mixed into the hydrogen-rich envelope.
Because neutrino loss and electron conduction  contribute very little
to the total energy loss in the following calculations, 
we can approximate total energy conservation by
\begin{equation}
L_{\rm ph}= L_{\rm nuc} + L_{\rm G},
\end{equation}
where $L_{\rm ph}$ is the photospheric luminosity, $L_{\rm G}$ is 
the total gravitational energy release rate calculated by using
\begin{equation}
L_{\rm G} = \int_0^M \epsilon_{\rm g} d M_r,
\end{equation}
and $L_{\rm nuc}$ is the total nuclear energy release rate calculated by
using\begin{equation}
L_{\rm nuc} = \int_0^M \epsilon_{\rm n} d M_r,
\end{equation}
with $\epsilon_{\rm n}$  the nuclear burning rate per unit mass.

For the initial WD model, we adopted an equilibrium model
\citep[i.e., ``the steady-state models'' of ][]{nom07},
in which an energy balance is already established
between heating by mass accretion
and nuclear energy generation and cooling by radiative transfer
and neutrino energy loss.  This is a good approximation
of the long time-averaged evolution of a mass accreting WD
and we do not need to calculate many (thousands) cycles of shell flashes
to relax thermal imbalance in the initial condition when we
start from a cold WD.  Starting from such an initial equilibrium state,
the nova cycle approaches a limit cycle only after a few to several cycles
\citep[see, e.g.,][for more detail]{kat14shn}.


\subsection{A Test: Comparison with Idan et al.'s result}
\label{section_comparison}
We first check our numerical code by calculating shell flashes with
the same parameters as those in Idan et al. (2013), i.e., 
the same WD mass ($M_{\rm WD}=1.0~M_\sun$) and 
mass accretion rate ($\dot M_{\rm acc}=1\times 10^{-6}~M_\sun$~yr$^{-1}$),
and a very similar manipulation of mass accretion (on and off).
This accretion rate is above the stability line.
We plot the last five cycles of our forced nova calculation
in Figure \ref{L.m10.1E-6.two}(S1)--(S3). 
The accretion phase is indicated by the short horizontal blue lines
labeled ``acc'' in Figure \ref{L.m10.1E-6.two}(S1).  
The recurrence period is 9.2 yr. No mass ejection occurs. 
To compare with Idan et al.'s Figure 1, we see that  
our photospheric luminosity, 
$L_{\rm ph}$, in Figure \ref{L.m10.1E-6.two}(S1), 
recurrence period, and no mass ejection well reproduced 
Idan et al.'s results.   
Thus, we confirm that Idan et al.'s calculation describes a forced nova.
Enlarging Figure \ref{L.m10.1E-6.two}(S1, S2) would reveals a very subtle kink 
in $L_{\rm ph}$ and $L_{\rm nuc}$ at the point of restarting accretion
at $\log L_{\rm ph}/L_\sun = 3.5$ $( \log L_{\rm nuc}/L_\sun \sim 3)$. 
This would probably correspond to small wiggles around 
$\log L/L_\sun \sim 3.3$ in Figure 1 of \citet{ida13}, 
although the authors did not mention the cause of them. 
This  further supports the fact that our manipulation of accretion 
to be similar to them. 

Figure \ref{L.m10.1E-6.two}(S2) shows the nuclear luminosity,
$L_{\rm nuc}$, and gravitational energy release, $L_{\rm G}$,
for the same model as in Figure \ref{L.m10.1E-6.two}(S1). 
In the very early phase of the nova outbursts, nuclear burning produces 
a large flux, but most of this is absorbed into the bottom region of the 
hydrogen-rich envelope and the photospheric luminosity does not exceed 
the Eddington limit, so  $L_{\rm ph}$ has a flat peak. 
The burning zone slightly expands to 
absorb the nuclear energy flux, which appears in the negative values of
$L_{\rm G}$, although such negative values of $L_{\rm G}$ are 
cut off by the lower bound of the figure. 
The absorbed energy is gradually released 
in the later phase (positive values of $L_{\rm G}$). 
This energy release is a main source for 
the photospheric luminosity $(L_{\rm G} > L_{\rm nuc} > 0)$  
in the interpulse (quiescent) phase.  
Thus, the WD is still bright ($L_{\rm ph} > 2000~L_\odot$)
in the interpulse phase. 

One of the important features of this forced nova is
that the hydrogen-rich envelope does not expand so much. 
Thus, the photospheric temperature does not decrease to \textless300,000 K. 
Figure \ref{L.m10.1E-6.two}(S3) shows the photospheric temperature, 
$T_{\rm ph}$, supersoft X-ray (0.2--1 keV) flux, and UV band (1120--2640 \AA)
flux corresponding to the {\it Swift} $UVW2$ band.
These luminosities are calculated by using a blackbody assumption from the
photospheric temperature, $T_{\rm ph}$, and luminosity, $L_{\rm ph}$. 
They are very faint in the optical band, unlike classical novae.
Thus, such objects are recognized, if they exist, 
as intermittent supersoft X-ray sources (SSSs).  However,
an anticorrelation between optical (high state) and supersoft X-ray 
(off state) in the intermittent SSSs, RX~J0513.9$-$6951 and V~Sge
\citep[e.g.,][]{hac03RXJ, hac03VSge},
cannot be explained by these forced novae.
Another important feature is the very high duty cycle (nuclear burning on
$/$ cycle duration $=0.7$), i.e., 6.4 yr (on) + 2.8 yr (off) $=$ 9.2 yr
(cycle duration).  
Such a high duty cycle is not consistent with those of recurrent novae
as discussed in more detail in Section \ref{sec_discussion}.

In connection to Figure \ref{dmdtdM}, Idan et al.'s forced nova has
a smaller ignition mass than the critical envelope mass, i.e., 
$M_{\rm ig} < M_{\rm env,cr}=M_{\rm env}$(B).  After the onset 
of a shell flash, it reaches somewhere between A and B and moves leftward. 
This is consistent with the result that the nova has no wind mass-loss.
When it reaches point A, nuclear burning extinguishes. 
The WD becomes dark and moves downward.

Recently, \citet{hil15} also claimed that they did not obtain
steady hydrogen burning. However, the light curves shown in Figures 
1 and 5 of Hillman et al. indicate those are force nova models. 
In these light curves there are small jumps in luminosity around 
$\log L/L_\sun \sim 3.3$, which are signatures of re-starting 
matter accretion as seen in our Figure \ref{L.m10.1E-6.two}(L1).  
That is, they seems to re-start accretion too late to obtain steady burning.


\begin{figure}
\epsscale{1.15}
\plotone{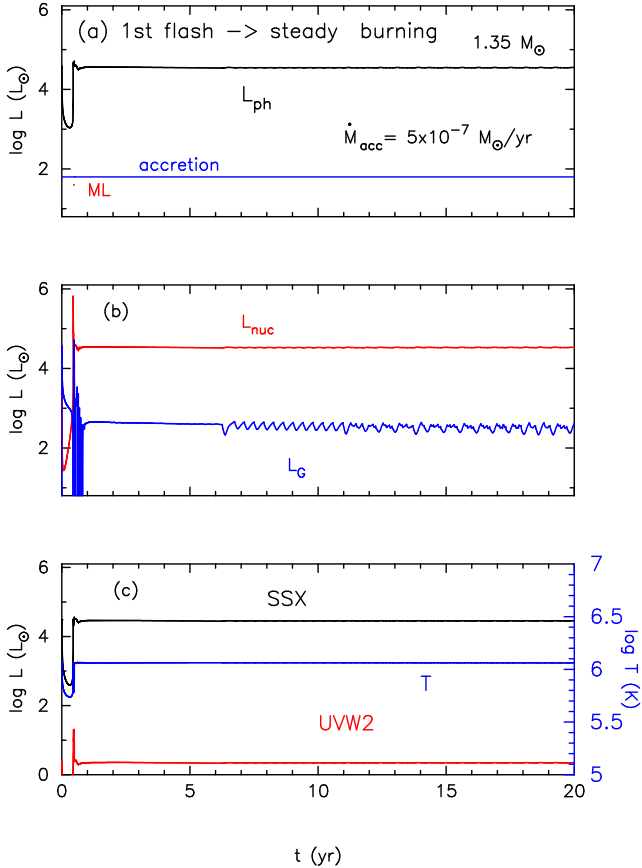}
\caption{
Same as  Figure \ref{L.m10.stburn}, but for a WD mass of $1.35~M_\sun$
and a mass accretion rate of $5\times 10^{-7}~M_\sun$~yr$^{-1}$.
The WD undergoes a first shell flash and
then settles down to  steady-state hydrogen shell burning.
\label{L.m135.stburn}}
\end{figure}

\subsection{First shell flash followed by steady hydrogen shell burning}
Figure \ref{L.m10.stburn} shows an example of our calculation of 
the first shell flash  on a $1.0~M_\sun$ WD with a mass accretion rate
of $\dot M_{\rm acc}=2.5\times 10^{-7}~M_\sun$~yr$^{-1}$,
which is located in the middle of the steady hydrogen-burning zone
in Figure \ref{PrecWDmass}  
($\dot M_{\rm stable} < \dot M_{\rm acc} < \dot M_{\rm cr}$).
This first shell flash occurs 20 yr after the start of mass accretion. 
The accreted matter is $\log M_{\rm env}~(M_\sun)\approx -5.3$, which
exceeds the critical envelope mass for wind mass loss, i.e., 
$M_{\rm env} > M_{\rm env,cr}=M_{\rm env}$(B),
in Figure \ref{dmdtdM}.  So, we have mass loss during
a short period (denoted by the orange line labeled ``ML''). 
Another example is an accreting 1.35 $M_\sun$ WD 
with $\dot M_{\rm acc}=5 \times 10^{-7}~M_\sun$~yr$^{-1}$, as
shown in Figure \ref{L.m135.stburn}
\citep[the same model as in Figure 7 of][]{kat14shn}.
It has a first shell flash 0.5 yr after the onset of accretion and maintains 
steady burning. 

In this way, the first shell flash always occurs if we start the mass
accretion onto a naked WD for any accretion rate above the stability line. 
Of course, this does not mean that the hydrogen shell burning is unstable.
We note that Starrfield et al.'s (2012) claim of the absence of steady
burning seems to be misguided from their calculations up to the first
flash; they didn't continue the calculation after the first flash.

In the models of Figures \ref{L.m10.stburn} and \ref{L.m135.stburn},
we resume mass accretion just after the wind mass loss stops,
i.e., before the hydrogen burning extinguishes
(before it reaches point A in Figure \ref{dmdtdM}).   
Hydrogen shell burning continues.
The accreted matter burns steadily at the same rate as the mass accretion.
The luminosity is constant with time in a balance of 
$L_{\rm ph}\approx L_{\rm nuc}$, 
and the envelope mass is also constant with time.  
Hydrogen shell burning is stable and never stops
as long as we maintain the mass accretion 
as indicated by the horizontal blue lines
in Figures \ref{L.m10.stburn}(a) and \ref{L.m135.stburn}(a).
A helium layer develops underneath the hydrogen burning shell.
A helium shell flash occurs when the mass of the helium layer
becomes large enough.  

We suppose that these binaries are in the steady state described in Figure 
\ref{steadyaccretion}.  The photospheric temperature of a WD is high
enough to emit supersoft X-rays.
Such an object could be recognized as a persistent luminous
supersoft X-ray source \citep[e.g.,][]{vdh92,kah97}.


\begin{figure*}
\epsscale{0.8}
\plotone{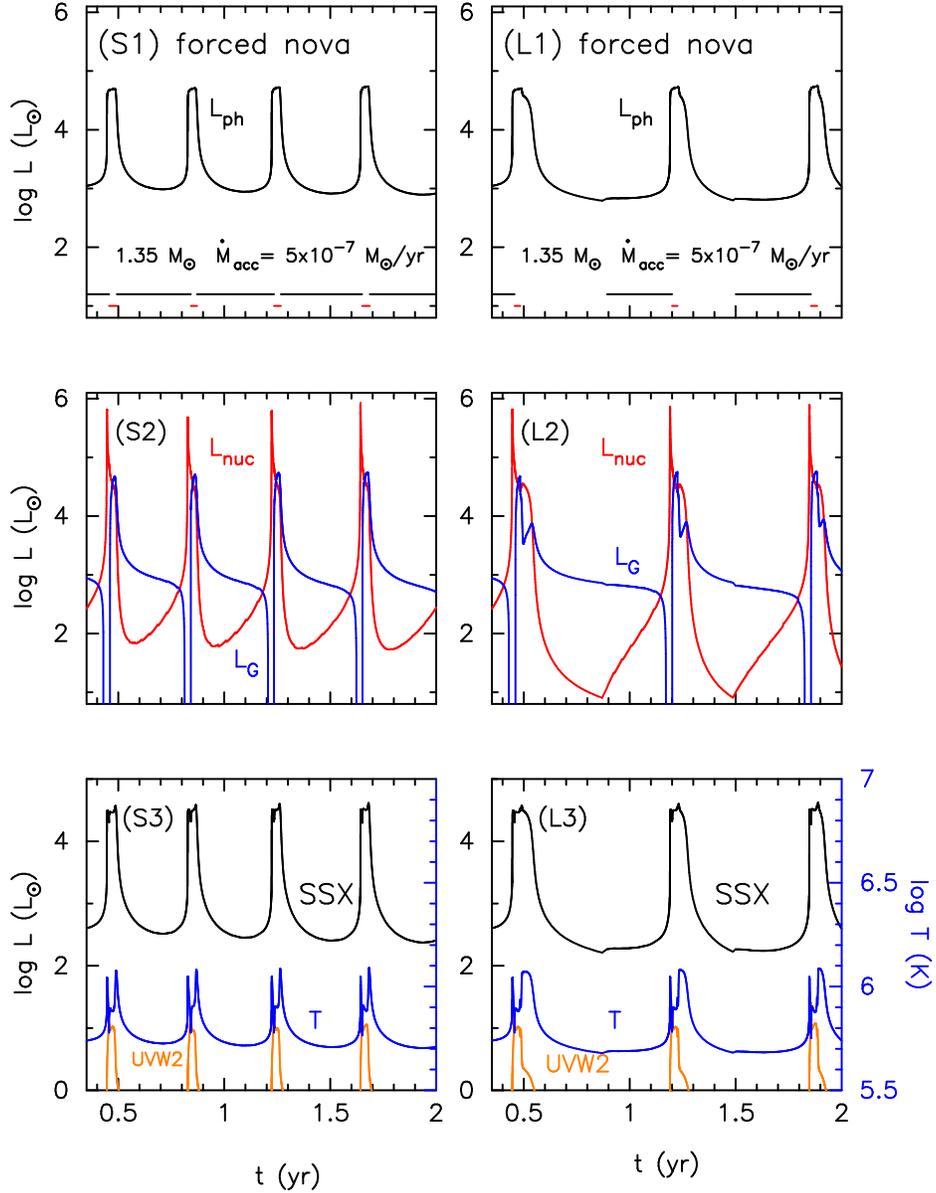}
\caption{
Shell flashes on a $ 1.35~M_\sun$ WD with 
a mass accretion rate of $5\times 10^{-7}~M_\sun$ yr$^{-1}$. 
If we resume the mass accretion immediately after hydrogen shell burning
ends, as indicated by the horizontal black lines, 
the next shell flash begins  every 0.42 yr in the 
left panels (S1--S3).
If we postpone the mass accretion until $L_{\rm nuc} < 10~L_\sun$, 
we get a longer recurrence period of 0.67 yr,
as in the right panels (L1--L3).
Horizontal black and red lines in panels (S1) and (L1) denote 
accretion phase and mass-loss phase, respectively. 
These two forced nova models are taken from \citet{kat14shn}.  
\label{L.m135.5E-7}}
\end{figure*}

\subsection{Forced novae: Manipulated mass accretion}
The on/off epochs of mass accretion can be designed to control
the recurrence period even for the same mass accretion rate and WD mass.
If we continue mass accretion during the bright $L_{\rm ph}$ phase
of the first shell flash, we obtain continuous hydrogen shell burning,
which never stops as long as we continue mass accretion
(until a helium shell flash occurs). 
However, if we stop at the beginning of a flash and resume mass accretion 
after hydrogen burning extinguishes, we obtain successive shell flashes. 
In this case, a later on time of mass accretion results in 
a longer recurrence period.

Figures \ref{L.m10.1E-6.two}(L1)--(L3)
demonstrate an example of 
designed mass accretion. For the same WD mass and mass accretion rate
as those of  Figures \ref{L.m10.1E-6.two}(S1)--(S3), 
we can get a longer recurrence period of 23.7 yr by omitting 
mass accretion for 7 yr. The outburst amplitude is larger 
than in the case of Figure \ref{L.m10.1E-6.two}(S1)
because of the cooling during the longer interpulse phase
(compare Figure \ref{L.m10.1E-6.two}(S2) 
with Figure \ref{L.m10.1E-6.two}(L2)). 
Since the ignition mass in Figure \ref{L.m10.1E-6.two}(L1)
is larger than that in Figure \ref{L.m10.1E-6.two}(S1),
the outburst duration is longer in Figure \ref{L.m10.1E-6.two}(L1).
The supersoft X-ray flux in the interpulse phase is lower 
in Figure \ref{L.m10.1E-6.two}(L3) than in 
Figures \ref{L.m10.1E-6.two}(S3). 
In this way, we can design the recurrence period
by changing the onset time of mass accretion.

Figures \ref{L.m135.5E-7}(S1)--(S3) and  (L1)--(L3)
depict other examples, that is, the same models as in Figures 7(b)
and 7(c), respectively, of \citet{kat14shn}, showing  
successive shell flashes for a forced nova on a $1.35~M_\sun$ WD
with a mass accretion rate of $5\times 10^{-7}~M_\sun$ yr$^{-1}$. 
In Figure \ref{L.m135.5E-7}(S1), we assume immediate resumption of mass 
accretion after hydrogen shell burning begins to decay,
as shown in the horizontal black (accretion) and red (mass loss) lines. 
If we restart the mass accretion a bit later, as shown 
in Figure \ref{L.m135.5E-7}(L1), we get a longer recurrence period. 
In this case, the nuclear burning region  cools much more than in the
case of Figure \ref{L.m135.5E-7}(S1),
so $\log L_{\rm nuc}~(L_\sun)$ dropped to 1.0 in the interpulse phase,
a tenth of that in the case of Figure \ref{L.m135.5E-7}(S2).
In contrast, the photospheric luminosity $\log L_{\rm ph}$
in Figure \ref{L.m135.5E-7}(L1) does not decrease
much in the interpulse phase but is almost the same as that 
in the case of Figure \ref{L.m135.5E-7}(S1), 
because it is supplied by the gravitational energy release $\log L_{\rm G}$.
The duration of outbursts is longer in Figure \ref{L.m135.5E-7}(L1)
than in Figure \ref{L.m135.5E-7}(S1) because the ignition mass 
is larger in Figure \ref{L.m135.5E-7}(L1)
owing to the cooling effect during the longer interpulse duration. 
In both cases, forced novae may be observed, if they exist, as an 
intermittent supersoft X-ray source.   However, the optical and
soft X-ray model light curves in Figures \ref{L.m10.1E-6.two} and
\ref{L.m135.5E-7} cannot explain the anticorrelation between the optical
high state and the supersoft X-ray off state observed
in RX~J0513.9$-$6951 and V~Sge.


\begin{figure}
\epsscale{1.15}
\plotone{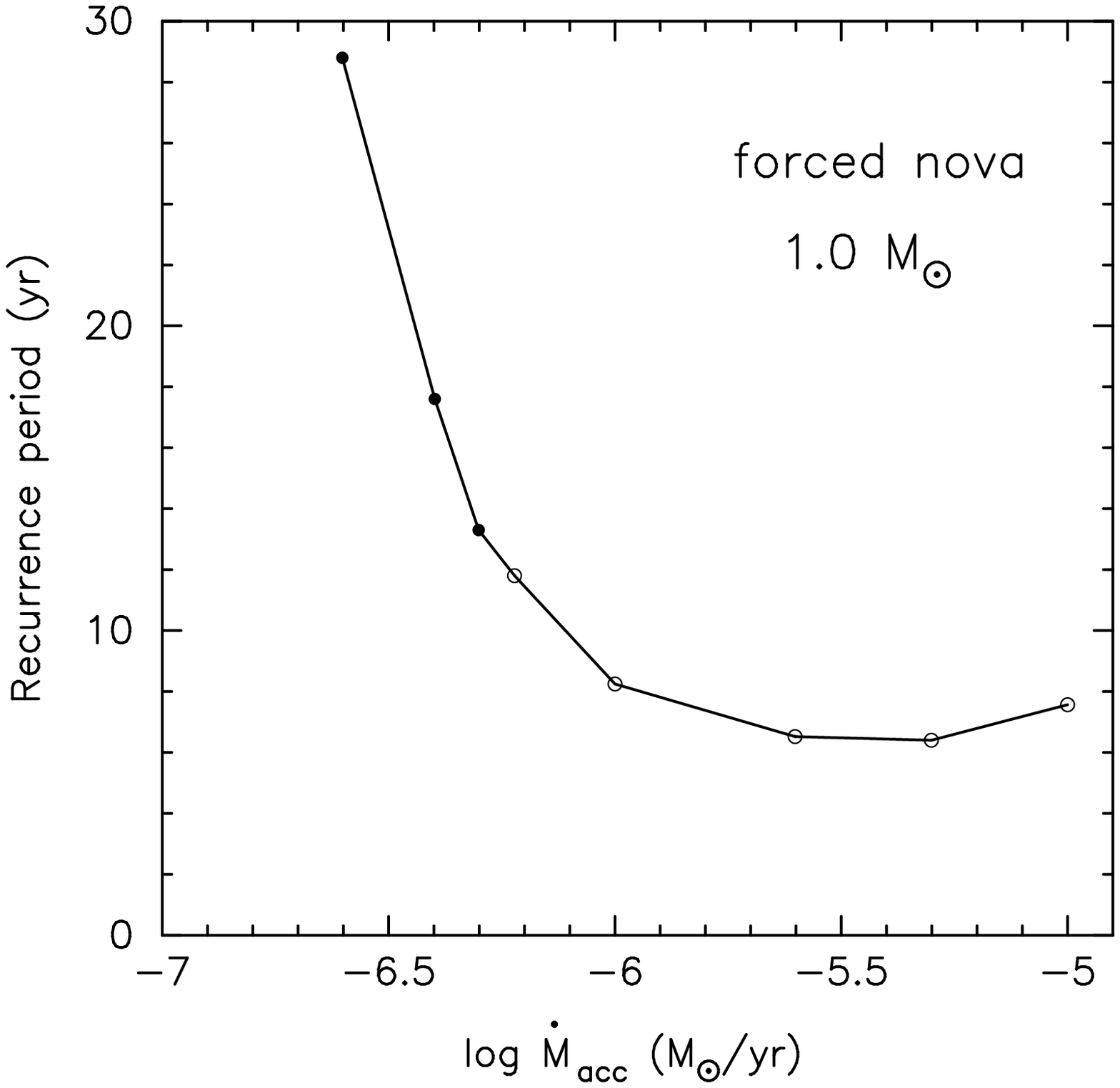}
\caption{Recurrence periods for various mass accretion rates on 
a 1.0 $M_\sun$ WD. Black dots indicate the case
in which mass loss occurs in the early phase of the shell flash,
while open circles indicate the case of no mass loss.
The minimum value is 6.4 yr
for $\dot M_{\rm acc}=5.0 \times 10^{-6}~M_\sun$~yr$^{-1}$.
\label{Prec.acc.m10}}
\end{figure}


\begin{figure}
\epsscale{1.15}
\plotone{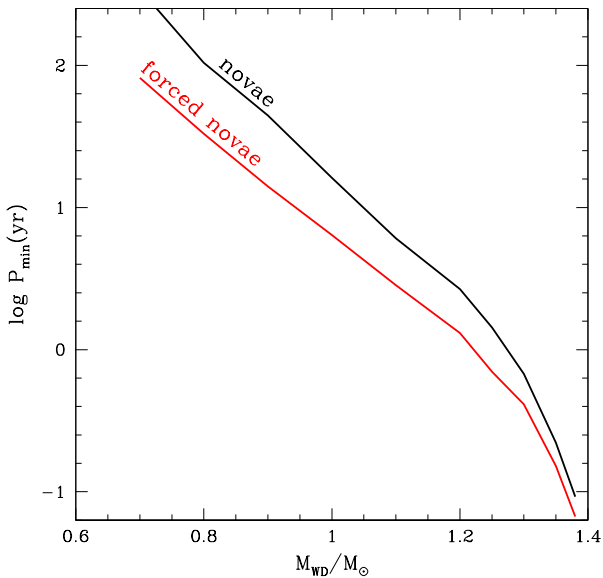}
\caption{Shortest recurrence periods for various WD masses. 
The red line indicates the minimum recurrence period of forced novae. 
The black line indicates the minimum recurrence period of natural novae,
which is taken from \citet{kat14shn}. See text for more details.
\label{Pmin}}
\end{figure}

\section{Shortest Recurrence Periods of Forced Novae}
\label{sec_Pmin}
Forced novae could have shorter recurrence periods than natural novae do.  
Because the recurrence period is a useful indicator of the WD mass of a 
recurrent nova, we obtain the minimum recurrence periods of forced novae. 

As shown in Figures \ref{L.m10.1E-6.two} and \ref{L.m135.5E-7},
we obtain the shortest recurrence period
for a given WD mass and mass accretion rate
if we switch on the mass accretion immediately after a shell flash ends
(i.e., when hydrogen shell burning begins to decay).
Here, we identify the end of the shell flash as 
when the photospheric luminosity decays
by a factor of 3 from that at the knee in the HR diagram. 
For a $1.0~M_\sun$ WD,  the switch-off luminosity is 
$\log L_{\rm ph}~(L_\sun) = 4.2$ 
and the switch-on luminosity is $\log L_{\rm ph}~(L_\sun) = 3.7$. 

Figure \ref{Prec.acc.m10} shows the recurrence periods of 
a $1.0~M_\sun$ WD for various mass accretion rates.  
For low accretion rates, the recurrence period decreases
as the accretion rate increases because the accretion time
to ignition mass becomes shorter.
For a very large mass accretion rate, the 
recurrence period increases because additional mass is accreted after the 
ignition (making the duration of hydrogen burning longer)
before the luminosity increases to the switch-off luminosity. 
We obtain the shortest recurrence period of 6.4 yr at 
$\dot M_{\rm acc}=5 \times 10^{-6}~M_\sun$~yr$^{-1}$
for a $1.0~M_\sun$ WD.  

The recurrence period of a nova, $t_{\rm rec}$,
is composed of 
\begin{equation}
t_{\rm rec} = t_{\rm nova-off} + t_{\rm nova-on},
\label{recurrence_time}
\end{equation}
where $t_{\rm nova-off}$ is the duration of the quiescent phase (accretion
phase, that is, hydrogen burning off) and $t_{\rm nova-on}$ is 
the duration of nova outburst (during which hydrogen shell
burning is occurring).  In classical novae, $t_{\rm nova-on}$ is much shorter
($\sim $$1$ yr or so) than $t_{\rm nova-off}$ ($10^3$--$10^5$ yr or so),
so we may neglect it.
Although neglecting $t_{\rm nova-on}$ is a very good approximation
for longer recurrence periods, slight deviations are appreciable
for recurrence periods of 1 and 3 yr in Figure 1. 

In this manner, we obtained the recurrence periods
of forced novae for various WD masses and mass accretion rates
and plotted them in the mass accretion rate versus WD mass diagram
in Figure \ref{PrecWDmass} by using red lines for $t_{\rm rec}=1$, 3, 10, 
and 30 yr. 
The red lines (forced novae) for recurrence periods of 1 and 3 yr shift
slightly from the corresponding black lines (natural novae)
at the stability line (dashed line). The shifts indicate
that the recurrence period of a forced nova for a given $M_{\rm WD}$
and $\dot M_{\rm acc}$ is slightly longer 
than that of the corresponding natural nova.
This comes from the fact that recurrence times of natural novae 
\citep[adopted from][]{kat14shn} were calculated  by neglecting the 
$t_{\rm nova-on}$ time, while recurrence periods of 
forced novae are obtained by calculating full cycles of novae. 

We summarize the shortest recurrence periods of forced novae 
for various WD masses as shown by the red solid line in Figure \ref{Pmin}. 
In the region below the red line, we do not have corresponding objects. 
In the region above the red line, forced novae can exist 
if we introduce some switch on/off mechanism.
In general, for a given WD mass, a longer recurrence period is
obtained if we interrupt the mass accretion for a longer time. 
Thus, the duty cycle is higher for a shorter recurrence period. 

The shortest recurrence periods of natural novae are obtained
near the stability line. We also plot the minimum value by the black
line in Figure \ref{Pmin}.  The values are taken from the recurrence period
on the line of $\dot M_{\rm stable}$ in Figure \ref{PrecWDmass}
\citep[with the original value being in Figure 6 of][]{kat14shn}.  
In the region below the black line, we have no natural novae. Above the line, 
shorter recurrence periods correspond to recurrent novae. The duty cycle is 
also higher for a shorter recurrence period. 
These black and red lines clearly show that
the minimum recurrence period of a forced nova
is always shorter than that of a natural nova.

\section{Discussion}
\label{sec_discussion}
Now, we compare our calculated recurrence periods
with those of recurrent novae 
\citep[see, e.g.,][for a review]{kat12review}. 

\subsection{T Pyx} \label{sec_TPyx} 
T~Pyx is a recurrent nova with six recorded outbursts in 1890,
1902, 1920, 1944, 1966, and 2011.  The orbital period of 1.83~hr was
obtained by \citet{uth10}. 
From the minimum periods of novae shown in Figure \ref{Pmin}, 
we estimate a lower limit of the WD mass to be $M_{\rm WD} > 0.93~M_\sun$ for
$t_{\rm rec}=12$~yr,
taking into account the possibility of it being a forced nova.
Note that WDs close to this lower limit mass exhibit a high 
duty cycle, which is inconsistent with the observed
low duty cycle of T~Pyx ($0.012 = 200$~days$/44$~yr). 
Thus, we may conclude that the WD mass of T~Pyx should be 
$M_{\rm WD} \gg 0.93~M_\sun$.

\citet{pat14} reported a stable increase of the orbital period 
of T~Pyx during 1966--2011, i.e., $P/\dot P \sim 3\times 10^5$~yr. 
This period change suggests a mass transfer
rate of $\dot M_{\rm acc}\sim 1\times 10^{-7}~M_\sun$~yr$^{-1}$
for $M_1=M_{\rm WD}\sim 1~M_\sun$ and a mass ratio of $q=M_2/M_1\sim 0.1$,
where $M_1$ and $M_2$ are the primary (WD, accretors) and secondary 
(main-sequence, donor) components, respectively.
This accretion rate combined with the WD mass ($M_{\rm WD}\gg0.93~M_\sun$) 
suggests that T Pyx is not a forced nova but rather is a natural nova because 
it is under the stability line in Figure \ref{PrecWDmass}.

Figure \ref{PrecWDmass} also suggests the WD mass to be 
$M_{\rm WD}\sim 1.1~M_\sun$ from the recurrence period of 
$t_{\rm rec} \sim 44$~yr during 1966--2011
with $\dot M_{\rm acc}\sim 1\times 10^{-7}~M_\sun$~yr$^{-1}$. 
This value is very consistent with 
the above lower limit mass of $M_{\rm WD}\gg0.93~M_\sun$.  
However, our estimate of the WD mass is inconsistent with
the estimate $M_{\rm WD}=0.7\pm0.2~M_\sun$ by 
\citet{uth10} based on the time-resolved spectroscopy of T~Pyx.

\subsection{Other recurrent novae} \label{sec_RN} 
The shortest record of recurrence periods is 1 yr of 
the recurrent nova M31N 2008-12a 
\citep[e.g.,][]{dar14, hen14, tan14, dar15, hen15a, hen15b}. 
We pose a constraint of the WD mass to be 
$ > $$1.25 ~M_\sun$ from Figure \ref{Pmin}.
However, a nova close to the minimum period of forced novae
(red line) should show
a very high duty cycle close to 1.0, which means it will 
always be bright in  the optical or X-ray region.
This is inconsistent with the observed short duration 
of M31N 2008-12a ($\sim$$0.05\approx$18~days/1~yr).  
This suggests that $ M_{\rm WD} \gg 1.25 ~M_\sun$.
\citet{kat15sh} obtained $M_{\rm WD}\sim 1.38~M_\sun$
from the duration of the supersoft X-ray phase and the mass accretion
rate of $\dot M_{\rm acc} = 1.6\times10^{-7}~M_\sun$~yr$^{-1}$.
The mass accretion rate is below the stability line, so this nova
should be a natural nova. 

The Galactic recurrent nova U~Sco exhibits recurrence periods
of $t_{\rm rec} \sim 8$--12~yr, which indicate that
the WD mass should be as massive as 
$M_{\rm WD} > 1.0~M_\sun$ based on Figure \ref{Pmin}.
\citet{hkkm00} modeled the light curve of U Sco and suggested the WD mass
to be $1.37~M_\sun$.  \cite{tho01} obtained a dynamical WD mass  of $M_{\rm WD}=1.55\pm 0.24~M_\sun$, which is consistent with
Hachisu et al.'s value.
The pair of $t_{\rm rec} \sim 8$--12~yr
and $M_{\rm WD}\sim 1.37~M_\sun$ is located above the minimum period
for natural novae (black line) in Figure \ref{Pmin}, indicating
U~Sco to be a natural nova.

CI~Aql is also a recurrent nova with  recorded outbursts in
1917, 1941, and 2000.  \citet{sch01c} proposed that CI~Aql could
outburst every $\sim$$20$~yr.  The recurrence period
of $t_{\rm rec} \sim 20$~yr requires the WD to be as massive as 
$M_{\rm WD} > 0.95~M_\sun$ (a natural nova) or
$M_{\rm WD} > 0.85~M_\sun$ (a forced nova) 
from Figure \ref{Pmin}.
\citet{sah13} estimated a dynamical WD mass of 
$1.0\pm0.14~M_\sun$ from their spectroscopic analysis.  
The pair of $t_{\rm rec} \sim 20$~yr
and $M_{\rm WD}\sim 1.0~M_\sun$ is located above but  close to 
the minimum period of natural novae (black line) in Figure \ref{Pmin}.
In such a case, the duty cycle (hydrogen burning on/cycle duration) of the model is rather high
and not consistent with the observation.
Therefore, we prefer a mass close to the upper bound 
$M_{\rm WD}\sim 1.14~M_\sun$ or higher.  This is consistent with
the mass $M_{\rm WD}=1.2\pm0.05~M_\sun$ estimated by
\citet{hac03ks} for CI~Aql from their light curve fitting.

Thus, we have no observational evidence of forced novae.  
There is, however, a possibility that forced novae will be discovered
in the future.  For example, we consider a wind-fed type
mass accretion, in which a WD accretes a part of cool wind from 
a red-giant companion after evacuated by the nova ejecta.
In the case of RS~Oph \citep{hac01kb},
which is not a forced nova but a natural nova,
the post-outburst minimum ($\sim 1$ mag fainter than the normal
quiescent phase) lasted about 300 days from $\sim 100$ to $\sim 400$ days
after the outburst \citep[see, e.g., Figure 2 of][]{hac06b}, suggesting  
the wind-fed accretion to the WD resumed 
much later than the end of hydrogen shell burning
$\sim 90$ days after the outburst \citep{hac06b, hac08kl}.
Thus, the mass accretion can be switched off for a long
time in some binary configuration.
If a similar system exists and its mass-accretion rate is higher
than the stability limit, we could have a forced nova.


\section{Conclusions}
\label{sec_conclusion}
Our main results are summarized as follows.

\begin{enumerate}
\item
We calculated nova outbursts using a Henyey-type evolution code 
with high mass accretion rates above the stability line, i.e.,
$\dot M_{\rm acc} > \dot M_{\rm stable}$. 
In our models, we assume that the accretion disk is not blown off
by optically thick winds (envelope mass ejection) and that the mass accretion
continues throughout the shell flash.  Then, we confirmed that steady
hydrogen shell burning is occurring in a zone above the stability line
of the WD mass versus mass accretion rate diagram.  
Thus, we obtained a first shell flash and subsequent steady hydrogen
shell burning for the mass accretion rate above the stability line.

\item 
If we stop mass accretion during the wind phase (mass ejection phase)
of nova outbursts, the resuming time of mass accretion controls
the subsequent shell flashes even for a mass accretion rate above
the stability line.  We named such shell flashes ``forced novae.''  
The recurrence periods of forced novae can be freely designed
by changing the resuming time of mass accretion. 
For a given WD mass and mass accretion rate, the shortest recurrence
period is obtained under the condition that we resume the mass accretion
immediately after the end of a shell flash. 
For a fixed WD mass, the shortest recurrence period thus obtained
attains a minimum at some mass accretion rate.  
The minimum recurrence periods are obtained for various WD masses.  

\item
We clarified the reason why in some recent works the claim has been made that  
nova outbursts have occurred above the stability line instead of steady-state burning.
These works used calculations with manipulated mass accretion; 
successive shell flashes are shaped by assuming a periodic
mass accretion.

\item
We constrain the WD mass of T Pyx to be $M_{\rm WD} > 1.0~M_\sun$
from its recurrence periods even for the case of forced novae.
This is not consistent with Uthas et al.'s (2010) claim of
$M_{\rm WD}=0.7\pm0.2~M_\sun$ based on the time-resolved spectroscopy.

\end{enumerate}

\acknowledgments
We thank the anonymous referee for useful comments that 
improved the manuscript.
  This research was supported in part by 
Grants-in-Aid for Scientific Research (24540227 and 15K05026)
from the Japan Society for the Promotion of Science.

%












\end{document}